\documentclass[aps,prx,onecolumn,superscriptaddress,longbibliography]{revtex4-2}

\newcommand{\bra}[1]{\mbox{$\langle#1|$}}
\newcommand{\ket}[1]{\mbox{$|#1\rangle$}}

\usepackage{graphicx} 

\usepackage{fullpage}
\usepackage{diagbox}

\usepackage{caption}

\usepackage{makecell}
\usepackage{multirow}
\usepackage{anyfontsize}

\usepackage{diagbox}

\usepackage[toc,page]{appendix}

\usepackage[american]{babel}

\usepackage{amsmath}
\usepackage{amssymb}
\usepackage{graphicx}
\usepackage[caption=false]{subfig}
\usepackage{color}
\usepackage[percent]{overpic}
\usepackage{bm}
\usepackage{rotating}

\usepackage{hyperref}

\usepackage{pbox}
\usepackage{array}
\usepackage{physics}
\usepackage{mathtools}
\usepackage{enumerate}
\usepackage{leftidx}
\usepackage{comment}
\usepackage{booktabs}
\usepackage{cancel}

\usepackage{xcolor}

\usepackage[normalem]{ulem}

\usepackage{float}

\usepackage{cleveref}
\usepackage{indentfirst}

\begin{document}

\title{{\it Reply to comment on} \\
Observation of the quantum equivalence principle for matter-waves}

\author{Or Dobkowski}
\affiliation{Ben-Gurion University of the Negev, Department of Physics and Ilse Katz Institute for Nanoscale Science and Technology, Be'er Sheva 84105, Israel}

\author{Barak Trok}
\affiliation{Ben-Gurion University of the Negev, Department of Physics and Ilse Katz Institute for Nanoscale Science and Technology, Be'er Sheva 84105, Israel}

\author{Peter Skakunenko}
\affiliation{Ben-Gurion University of the Negev, Department of Physics and Ilse Katz Institute for Nanoscale Science and Technology, Be'er Sheva 84105, Israel}

\author{Yonathan Japha}
\affiliation{Ben-Gurion University of the Negev, Department of Physics and Ilse Katz Institute for Nanoscale Science and Technology, Be'er Sheva 84105, Israel}

\author{David Groswasser}
\affiliation{Ben-Gurion University of the Negev, Department of Physics and Ilse Katz Institute for Nanoscale Science and Technology, Be'er Sheva 84105, Israel}

\author{Maxim Efremov}
\affiliation{German Aerospace Center (DLR), Institute of Quantum Technologies, 89081 Ulm, Germany}
\affiliation{Institut für Quantenphysik and Center for Integrated Quantum Science and Technology ($\it IQST$), Universität Ulm, 89081 Ulm, Germany}

\author{Chiara Marletto}
\affiliation{Clarendon Laboratory, University of Oxford, Parks Road, Oxford OX1 3PU, United Kingdom}

\author{Ivette Fuentes}
\affiliation{School of Physics and Astronomy, University of Southampton, Southampton SO17 1BJ, United Kingdom}
\affiliation{Keble College, University of Oxford, Oxford OX1 3PG, United Kingdom}

\author{Roger Penrose}
\affiliation{Mathematical Institute, Andrew Wiles Building, University of Oxford, Radcliffe Observatory Quarter, Woodstock Road, Oxford, OX2 6GG, United Kingdom}

\author{Vlatko Vedral}
\affiliation{Clarendon Laboratory, University of Oxford, Parks Road, Oxford OX1 3PU, United Kingdom}

\author{Wolfgang P. Schleich}
\affiliation{Institut für Quantenphysik and Center for Integrated Quantum Science and Technology ($\it IQST$), Universität Ulm, 89081 Ulm, Germany}
\affiliation{Hagler Institute for Advanced Study at Texas A\&M University, Texas A\&M AgriLife Research, Institute for Quantum Science and Engineering (IQSE), and Department of Physics and Astronomy, Texas A\&M University, College Station, Texas 77843-4242, USA}

\author{Ron Folman}
\affiliation{Ben-Gurion University of the Negev, Department of Physics and Ilse Katz Institute for Nanoscale Science and Technology, Be'er Sheva 84105, Israel}

\date{\today}

\begin{abstract}
We show that in contrast to a recent claim, the Quantum Galileo Interferometer is sensitive to a uniform gravitational field in the presence and even in the absence of the levitation condition.
\end{abstract}


\maketitle

\section{In a nutshell}

In a recent article \cite{QGI} we have reported the experimental realization of the Quantum Galileo Interferometer (QGI), which is sensitive to the square of the gravitational acceleration and the cube of time that the atom spends in a uniform gravitational field. The authors of a recent comment \cite{QGI-comment} claim that the measured phase shift in the QGI does not result from gravity but rather from the magnetic field gradient used to apply a force on the atom in one branch of the QGI. Specifically, they show that the phase can be calculated just from the magnetic gradient without any need for any knowledge of gravity or the acceleration $g$ due to gravity, and hence, so they claim, the QGI does not measure gravity. 

In the present reply we demonstrate that this statement is wrong and results from a misunderstanding of the conditions required to operate the interferometer. Specifically we show that for the QGI interferometer to work one must apply a ``levitation condition", and this necessarily introduces $g$ as a crucial parameter. We then generalize this result, and show that even in the absence of the levitation condition, but applying a more general ``closing condition" so that the interferometer loop is closed in space-time, the QGI is still sensitive to $g$. 

Generally speaking, we believe that the authors of the comment were misled by their incorrect assertion that ``a uniform gravitational field is unobservable", which originates from a misinterpretation of the meaning of the Einstein Equivalence Principle (EEP). In contrast to their assertion, the EEP does indeed allow to measure (observe) gravity, if one has a reference frame external to the system, typically the lab frame. It is for this reason that ever since Galileo and Newton and to this day, scientists have been able to measure gravity with a wide range of tools.

\section{Levitation and closing conditions}

We first briefly summarize the essential ingredients of the QGI and then calculate the interferometer phase both by correctly calculating the actions and by utilizing Feynman's propagator technique \cite{T3}. The latter representation-free approach \cite{Schleich-NJP, Schleich-PRL} avoids ambiguities in the interpretation of the origin of the phase shift arising usually within the semi-classical approach based on actions. We also present the general arguments in favor of being able to measure a uniform gravitational field when employing the QGI.

In the QGI \cite{QGI}, we compare the phase experienced by an atom in free-fall with an atom at rest. Generally speaking, the resulting phase shift arises from the Kennard phase \cite{T3,T3-PRL}, which is the phase between two interferometric wave packets experiencing different accelerations.

The two arms of the interferometer are formed by the center-of-mass motion of two internal states of an atom. As in a Ramsey interferometer, a microwave pulse prepares a superposition of the two states and mixes them again after a time of evolution. During a time $T$ the internal states feel different accelerations and therefore accumulate, during their center-of-mass motions, two different phases. In the laboratory frame, the atom in state $|F,m_F\rangle=|2,1\rangle$, called the reference wave packet, is levitated by a magnetic field gradient $\nabla B_{lev}$ providing a force which is opposite to the gravitational force of Earth. The atom in state $|1,0\rangle$, called the ballistic wave packet, is insensitive to the magnetic field gradient and only experiences the gravitational potential. To start the motion in this state, a catapulting magnetic gradient pulse $\nabla B_{cat}$, generally independent of the previous magnetic field gradient, imparts a momentum $p_0 = m_i v_0$ at $t = 0$ and
again at $T$ (to enable this, a microwave $\pi$ pulse briefly allows the exchange of states, see [1]). In the following, we keep track of the appearance of the inertial and gravitational masses, $m_i$ and $m_g$, just as the authors of the comment did, so that the exact differences in the two calculations are easily understood.

Before we address the points raised by the comment, we emphasize that two essential conditions ensure the appearance of the gravitational field in the phase shift of the QGI: (i) the levitation condition, $m_i a=m_g g$, which connects the acceleration $a$, caused in our experiment by the magnetic field gradient $\nabla B_{lev}$, with the uniform gravitational field and the two different types of masses, where $a=\mu \nabla B_{lev}/m_i$, and (ii) the closing condition connecting the momentum transfer due to gradient $\nabla B_{cat}$ with the interferometer time, namely $2v_0=aT$, which guarantees the closing of the interferometer in position and momentum.

In their comment \cite{QGI-comment}, the authors have assumed the closing condition. However, crucially (see also Discussion), they did not explicitly write down the levitation condition, although their final result [Eq.\,(6)] is only valid if this condition applies. Applying this levitation condition to their final result shows that gravity determines the phase. We also briefly note, that by their own admission, the authors chose to ignore several terms in the calculation of the action, leading to additional confusion. For completeness, we provide in Appendix A the full required calculation in the notation of the comment, including the magnetic force in the action of the reference wave packet and the action of the catapulting magnetic pulses, which were ignored.

In addition, it is important to explain that magnetic fields play no unique role in this discussion. To realize a QGI, the mechanisms of levitation and catapulting are independent of each other and two different fields, not necessarily magnetic ones, can create them. For example, the levitation could be optical while the catapulting at the beginning and the end could originate from an electric field.

Finally, we note that if the levitation condition is not executed perfectly, a residual acceleration $\Delta a$ exists for the reference wave packet, resulting in a phase term dependent on $g$. As detailed in the Discussion, this is also ignored in the main analysis provided in the comment \cite{QGI-comment}.

We now employ the representation-free approach \cite{Schleich-NJP,Schleich-PRL}, which, as proven in previous confusions, is able to award us with an unbiased view of the physics. To generalize our claim that the QGI measures gravity, we show that even in the absence of the levitation condition, the phase shift of the QGI is governed by the gravitational acceleration, provided the closing condition is applied appropriately. We perform the quantum calculation based on the relevant propagators \cite{T3}.

\section{Interferometer phase without levitation condition}

In the present discussion we allow for a non-zero residual acceleration $\Delta a\equiv a -(m_g/m_i) g$ of the reference wave packet. Due to this acceleration the reference wave packet starting with zero velocity from the origin has moved to the position $\Delta aT^2/2$ after the time $T$. For a given initial velocity $v_0=\nabla B_{cat}\tau/m_i$ (where $\tau$ is the magnetic pulse time), the ballistic wave packet, also starting from the origin, has moved in the time $T$ to the position $v_0T-(m_g/m_i)gT^2/2$. In order to close the interferometer in space, we have to choose the time $T=2v_0/[(m_g/m_i)g+\Delta a]$, or equivalently, for a specific $T$ set $v_0$ accordingly. The difference $\Delta a T-v_0+(m_g/m_i)gT=v_0$ of the velocities of the two wave packets at the time $T$ is compensated by the second closing pulse. We emphasize that even in the absence of the levitation condition, the two catapult pulses transfer the same velocity $v_0$. This feature is also confirmed by our quantum-mechanical consideration.

Indeed, we now present the quantum-mechanical representation-free treatment \cite{Schleich-NJP,Schleich-PRL} of the interferometer phase. We recall \cite{Schleich-NJP, Schleich-PRL} that the probability $P$ to count atoms in one of the exits of an atom interferometer reads
\begin{align}
    P = \frac{1}{4} \left( 2 + I + I^*\right)
    \label{Eq:CountProb}
\end{align}
with the interference term
\begin{align}
    I\equiv\bra{\psi}  \hat{U}_1^\dagger \hat{U}_2^{\vphantom{\dagger}} \ket{\psi}.
    \label{Eq:InterferenceTerm}
\end{align}
Here $\ket{\psi}$ denotes the initial state of the center-of-mass motion and $\hat{U}_1$ and $\hat{U}_2$ are the unitary time evolution operators corresponding to two branches $1$ and $2$ of the interferometer.

For the QGI, the expectation value $I$ takes the form
\begin{align}
    I &\equiv \langle \psi | e^{\frac{i}{\hbar} \hat{H}_{\Delta a} T} e^{\frac{i p_0 \hat{z}}{\hbar}} e^{-\frac{i}{\hbar} \hat{H}_g T} e^{\frac{i p_0 \hat{z}}{\hbar}} | \psi\rangle
\end{align}
with the Hamiltonians
\begin{align}
    \hat{H}_{\Delta a} &\equiv \frac{\hat{p}^2}{2m_i}- m_i \Delta a\hat{z},\;\text{and}\; \hat{H}_g \equiv \frac{\hat{p}^2}{2m_i} + m_g g\hat{z}.
\end{align}

Moving from the right to the left, the four unitary operators correspond to the momentum transfer by an amount of $p_0$ due to the first pulse, motion in the uniform gravitational field according to $\hat{H}_g$, and the momentum transfer $p_0$ by the second pulse. These operators act on the ballistic wave packet. The final operator involving $\hat{H}_{\Delta a}$ describes the motion of the reference wave packet with the constant residual acceleration $\Delta a$, however going backwards in time due to the Hermitian conjugate of $\hat{U}_1$.

In Appendix B we evaluate this expectation value and find the expression
\begin{align}
    I =\exp\left[-\frac{i}{24\hbar}\left(\frac{m_g^2 g^2}{m_i}- m_i(\Delta a)^2\right)T^3\right].
    \label{I-result}
\end{align}

The fact that the interference term $I$ is a phase factor with $|I|=1$ indicates that the QGI is not only closed according to classical mechanics, that is, in the coordinates and momenta given by the classical trajectories, but also according to quantum theory. The reason for this fact is that the spreading of a wave packet in a linear potential is independent of its slope, and entirely governed by the free motion, being identical for the reference and the ballistic wave packets.

According to Eq.\,\eqref{I-result}, the interference term contains the difference of the Kennard phases due to a uniform gravitational field and the residual acceleration, and demonstrates that even without the levitation condition, but with an appropriately adjusted interferometer time to achieve closing of the interferometer loop, a uniform gravitational field enters into the phase of the QGI. Moreover, the inertial and gravitational masses both appear in a non-trivial way in the gravitational Kennard phase. In the presence of the levitation condition, that is $\Delta a = 0$, we obtain the results of Ref. \cite{QGI}.

Since $\Delta a$ also contains $g$, it is instructive to cast $I$, Eq.\,\eqref{I-result}, into the form
\begin{align}
    I =\exp\left[\frac{i}{24\hbar}\left(1-2\frac{m_g}{m_i}\frac{g}{a}\right)m_i a^2T^3\right],
    \label{I-result2}
\end{align}

which demonstrates clearly that $I$ is sensitive to $g$. In the presence of the levitation condition this expression reduces to our result \cite{QGI} (if one takes the interferometer time to be $2T$, as is done in our paper \cite{QGI} and in the comment \cite{QGI-comment}, the prefactor $1/24$ turns into $1/3$).

We emphasize that the QGI is another realization of a $T^3$-atom interferometer \cite{T3,T3-Ron-PRL} and measures the difference of the Kennard phases caused by the two different accelerations in the two arms of the interferometer. In the presence of the levitation condition, the acceleration of the ballistic wave packet is $-g$, whereas the acceleration of the reference wave packet vanishes. Hence, the total phase difference is solely given by the Kennard phase corresponding to the uniform gravitational field.


We note again that our arguments hold true for any types of forces used for the momentum kicks and the levitation, e.g., optical, electric, magnetic, etc.

\section{Discussion}

We now conclude with a summary, including more general arguments against the claim that a uniform gravitation field is unobservable. We start with physics nomenclature considerations and end with a summary of our rigorous scientific arguments.

If one claims that what is measured in our experiment is in reality the magnetic force holding the reference wave packet static relative to Earth, then in perfect analogy, it should be justified to conclude that when Newton was sitting on Earth and an apple fell -- constituting for him a measurement of gravity -- the correct interpretation of this would be that all that he observed was the force of the ground atoms holding him static relative to Earth.

We believe that the authors of the comment were misled by their incorrect assertion that ``a uniform gravitational field is unobservable", which originates from a misinterpretation of the meaning of the Einstein Equivalence Principle (EEP). In contrast to their assertion, the EEP does indeed allow to measure (observe) gravity, if one has a reference frame external to the system, typically the lab frame. This is why ever since Galileo and Newton and to this day, scientists have been able to measure gravity with a wide range of tools.

As defined in Ref.\,[2] of the comment \cite{QGI-comment}, the exact wording of the EEP is: ``Fundamental non-gravitational test physics is not affected, locally and at any point of spacetime, by the presence of a gravitational field.”
What is meant by ``locally” is that all physics happens within a {\it closed} system, namely, there is no external reference frame or observer. However, in the multitude of experiments measuring gravity, the lab frame is typically an external reference frame. This feature is also true in our experiment. Indeed, the reference wave packet (from which we observe) is external to the freely-falling system.

Nowhere clearer than in the QGI do we recognize the importance of a "ruler" in the measurement of $g$ in an atom interferometer. Indeed, the phase accumulated by the ballistic wave packet during its travel in the uniform gravitational field, consisting of the Kennard phase and the one due to free wave packet evolution, is compared to the phase accumulated by the reference wave packet that is at rest with respect to the ballistic one. Hence, the reference wave packet only picks up a phase due to free wave packet evolution which is identical to the one of the ballistic wave packet. As a result, the interferometer phase is the Kennard phase due to gravity.


We now summarize our rigorous scientific arguments:
For the most elementary understanding of the error in the comment \cite{QGI-comment}, one may turn to the final equation of the comment, Eq.\,(6), which in essence contradicts the entire message that gravity is unobservable. The reason is that Eq.\,(6), which coincides with the interferometer phase only if the levitation condition holds, has in it the magnetic gradient that was used to hold static the reference wave packet $\nabla B_{lev}$. If one introduces the levitation condition, the magnetic gradient itself is an accurate measure of the gravitational force $m_g g$. If $g$ would be different, the magnetic gradient $\nabla B_{lev}$ would have to be changed accordingly so as to maintain the levitation condition of the interferometer. In other words, the levitation condition creates an exact measure of the gravitational field. Specifically, the levitation condition is in fact the weighing of the atoms, which forces us to replace the magnetic force in  Eq.\,(6) of the comment \cite{QGI-comment} with $m_g g$.
If one insists on presenting only the magnetic gradient, as the authors of the comment did, they should have emphasized that this represents a special case in which $m_g g = \mu \nabla B_{lev}$ (the levitation condition), highlighting the dependence of the phase on $g$.

More so, presenting the phase as purely magnetic, as the authors of the comment did, does not represent the underlying physics of the general case of a closed $T^3$ interferometer in which $g$ cannot be factored out in any way. The authors of the comment tried towards the end of their comment (just before their concluding remarks) to rectify this by briefly speaking of a ``residual acceleration", $a_{res}$, without realizing or informing the reader that $a_{res}=(\mu \nabla B_{lev}-m_g g)/m_i $, thus introducing $g$ into the phase.

In our analysis we have employed the representation-free approach \cite{Schleich-NJP,Schleich-PRL}, which, as proven in previous confusions, is able to award us with an unbiased view of the physics. Our calculation clearly shows [Eq.\,\eqref{I-result2}] that $g$ cannot be factored out in the general case of a closed interferometer (before applying the levitation condition), and hence the QGI undoubtedly measures gravity.

As noted above, we believe that in this case the confusion originates from a misinterpretation of the EEP, which leads to the misconception that ``a uniform gravitational field is unobservable". Because we are observing from an external reference frame, we are able to measure gravity.

\begin{appendix}

\section{Action calculation}
In the comment, the authors calculate the phase of the QGI, using the action, while choosing to ignore ``inertial phase shifts". The calculation leads to confusion as it either leaves out important phase terms or implicitly assumes that the levitation condition applies. Here, we perform the calculation again, using the same notation as in the comment, but include all the phase terms, and show that the phase does depend on $g$, both in the general case of a closed $T^3$ interferometer and the special case that includes the levitation condition.

The ballistic wave packet has initial velocity $v_0$, and is exposed to the gravitational field. As in the comment, we use the notation $g_0$ for the gravitational field, such that the gravitational force is $-m_g g_0$.

\begin{equation}
    S_{\text {ballistic }} = \int_0^{2 T}\left[\frac{m_i}{2}\left(v_0-\frac{m_g}{m_i} g_0 t\right)^2-m_g g_0\left(v_0 t-\frac{m_g g_0}{2 m_i} t^2\right)\right] d t ,
\end{equation}
which yields
\begin{equation}
    S_{\text {ballistic }} =  \frac{8 (m_g g_0)^{2} T^{3}}{3 m_{i}} - 4 (m_g g_0) T^{2} v_{0} + T m_{i} v_{0}^{2} .
\end{equation}

But, to get the correct phase of the ballistic wave packet, we must calculate also the action during the catapulting pulses, during which there is a magnetic potential acting on the wave packet, resulting in an acceleration $a_{kick} = \mu\nabla B_{\text {pulse }}/m_i$ for a duration $\tau$, such that $a_{kick}= \frac{v_0}{\tau}$. The action during the first pulse is given by
\begin{equation}
    S_{\text {kick}}=\lim _{\tau \rightarrow 0}\int_{0}^{\tau}\left[\frac{m_i}{2}\left(-\frac{m_g}{m_i} g_0 t' +\frac{v_0}{\tau} t'\right)^2-(m_g g_0 - m_i \frac{v_0}{\tau})\left(-\frac{m_g g_0}{2 m_i} t'^2 +\frac{\frac{v_0}{\tau}}{2} t'^2\right)\right] d t'
\end{equation}
In the limit of $\tau \to 0$, the action during the first pulse is zero. For the second pulse, the initial velocity is $v_0 - \frac{m_g}{m_i} g_0 2T$, and the initial position is $-\frac{m_g g_0}{2 m_i} (2T)^2 +v_0\cdot 2T$. The action during the second pulse is given by
\begin{align}
    \lim _{\tau \rightarrow 0}S_{\text {kick}} = \int_{0}^{\tau} & \Bigg[\frac{m_i}{2}\left(v_0 - \frac{m_g}{m_i} g_0(2T+t') +\frac{v_0}{\tau} t'\right)^2 \nonumber \\
    & -(m_g g_0 - m_i \frac{v_0}{\tau})\left(-\frac{m_g g_0}{2 m_i} [(2T)^2+t'^2] +v_0\cdot 2T +\frac{\frac{v_0}{\tau}}{2} t'^2\right)\Bigg] d t'
\end{align}
which yields
\begin{equation}
    \phi_{\text {kick}} = \frac{2 m_i v_0 T}{\hbar}\left(v_0-\frac{m_g}{m_i} g_0 T\right)
\end{equation}

Inserting the closing condition $v_0=F_{m a g} / m_i \cdot T$ we get
\begin{equation}
    \phi_{\text {ballistic }}+\phi_{\text {kick }}=\frac{m_i}{\hbar}\left[\frac{8}{3} \bar{g}^2-6 a \bar{g}+3 a^2\right] T^3
    \end{equation}
where we have defined $\bar{g} = \frac{m_g g_0}{m_i}$ and $a=F_{\text {mag}}/m_i$.

The reference wave packet is initially at rest and is exposed to the gravitational field and a magnetic force $F_{\text {mag }} =\mu \nabla B$. The action of the reference wave packet is given by

\begin{equation}
    S_{\text {reference }}=\int_0^{2 T}\left[\frac{m_i}{2}\left(\frac{F_{\text {mag }}}{m_i} t-\frac{m_g}{m_i} g_0 t\right)^2-(m_g g_0 -F_{\text {mag}}) \left(\frac{F_{\text {mag }}}{2 m_i} t^2-\frac{m_g g_0}{2 m_i} t^2\right)\right] d t .
\end{equation}
Here, we have included the magnetic potential acting on the reference wave packet, which the authors of the comment have chosen to ignore in the main calculation. This yields

\begin{equation}
    \phi_{\text {reference}} =  \frac{1}{\hbar}T^{3} \cdot \left(\frac{8 (m_g g_0)^{2}}{3 m_{i}} - \frac{16 (m_g g_0) F_{\text {mag}}}{3 m_{i}} + \frac{8 F_{\text {mag}}^{2}}{3 m_{i}}\right)
\end{equation}

The total phase difference is given by the difference of the two interferometer arms, which yields

\begin{equation}
    \Delta \phi = \phi_{\text {ballistic}}+ \phi_{\text {kick}} - \phi_{\text {reference }} = -\frac{F_{\text {mag}}T^3}{3 m_i\hbar}[F_{\text {mag}}-2 m_g g_0],
\end{equation}
or in terms of the accelerations this can be written as
\begin{equation}\label{phase_result}
    \Delta \phi = \frac{m_i a T^3}{3 \hbar}[a-2 \bar{g}]
\end{equation}

This result shows that the phase difference does depend on the gravitational field. 

Next, we can introduce the levitation condition, which is given by $F_{\text {mag }} = m_g g_0$. Inserting the levitation condition yields the following phase difference

\begin{equation}
    \Delta \phi = - \frac{(m_g g_0)^{2} T^{3}}{3 m_{i}},
\end{equation}
which demonstrates the dependence of the phase difference on the gravitational field. 

Using $F_{\text {mag }} = m_g g_0$ this can also be written as
\begin{equation}
    \Delta \phi = - \frac{F_{\text {mag }}^{2} T^{3}}{3 m_{i}}.
\end{equation}

This form of the phase holds true only when the levitation condition is applied.

The authors of the comment have chosen to ignore the magnetic potential in the action calculation as well as the action due to the magnetic catapulting pulse, leading to confusion. Due to this, the authors state that the phase difference is independent of the gravitational field, which does not hold in the full calculation. The phase difference depends on the gravitational field, and the authors have chosen to ignore this in their calculation, either by omitting the magnetic potential, or by implicitly assuming that the levitation condition applies without writing it as part of the calculation. The authors briefly mentioned a ``residual acceleration" that results in an ``inertial phase shift". Let us complete this analysis by rigorously introducing the residual acceleration notation, such that $\Delta a = a - m_g g_0/m_i$, and insert it into Eq.\,\ref{phase_result}. This yields
\begin{equation}
    \Delta\phi = -\frac{T^3}{3 \hbar}\left(\frac{m_g^2}{m_i} g^2-m_i(\Delta a)^2\right) ,
\end{equation}
which again emphasizes the dependence of the phase on the gravitational field. The result of this appendix is consistent with the result derived in the main text via the representation-free approach.

\section{Interference term}

In this Appendix we evaluate the interference term
\begin{align}
    I &\equiv \langle \psi | e^{\frac{i}{\hbar} \hat{H}_{\Delta a} T} e^{\frac{i p_0 \hat{z}}{\hbar}} e^{-\frac{i}{\hbar} \hat{H}_g T} e^{\frac{i p_0 \hat{z}}{\hbar}} | \psi \rangle
\end{align}
by inserting three complete sets of position eigenstates and arrive at the expression
\begin{align}
    I =\int dz'' \int dz' \int dz\, \psi^*(z'') \langle z'' | e^{\frac{i}{\hbar} \hat{H}_{\Delta a} T} | z' \rangle e^{\frac{i p_0 z'}{\hbar}} \langle z' | e^{-\frac{i}{\hbar} \hat{H}_g T} | z \rangle e^{\frac{i p_0 z}{\hbar}} \psi(z),
    \label{Eq:I2}
\end{align}
with $ \psi(z)\equiv \langle z | \psi \rangle$ being the initial wave function in the position representation. Throughout this appendix the integrations over the coordinates extend from $-\infty$ to $\infty$.

We recall the propagators \cite{T3}
\begin{equation}
     \langle z'' | e^{\frac{i}{\hbar} \hat{H}_{\Delta a} T} | z' \rangle =N^* \exp \left[ -\frac{i}{\hbar} \frac{m_i}{2} \frac{(z'' - z')^2}{T} -\frac{i m_i\Delta a}{2 \hbar} (z'' + z')  T\right]e^{-i \phi_{\Delta a}},
     \label{Eq:Ga}
\end{equation}
and
\begin{equation}
    \langle z' | e^{-\frac{i}{\hbar} \hat{H}_g T} | z \rangle = N\exp \left[\frac{i}{\hbar} \frac{m_i}{2} \frac{(z' - z)^2}{T} - \frac{i m_g g}{2 \hbar} (z' + z)  T \right] e^{i \phi_g},
    \label{Eq:Gg}
\end{equation}
of the reference and the ballistic wave packet corresponding to a non-relativistic particle in a linear potential
with the normalization constant
\begin{align}
    N = \sqrt{\frac{m_i}{2 i \pi \hbar T}}
    \label{Eq:N}
\end{align}
and the two position-independent Kennard phases
\begin{align}
    \phi_{\Delta a} \equiv - \frac{1}{24} \frac{m_i(\Delta a)^2}{\hbar} T^3,\;\;\;\phi_{g} \equiv - \frac{1}{24} \frac{m_g^2 g^2}{\hbar m_i} T^3.
\end{align}

Inserting the two propagators into Eq.~\eqref{Eq:I2} yields
\begin{align}
\begin{split}
    I = e^{i \left(\phi_g -\phi_{\Delta a}\right)} |N|^2 \int dz'' \!\! \int dz' \!\!\int dz \,\, &\psi^*(z'') \psi(z) \exp \left\{ - \frac{i}{\hbar} \frac{m_i}{2T} \left[ (z'' - z')^2 - (z' - z)^2 \right]
    \right\} \\
    &\times \exp \left\{\frac{i}{\hbar}\left[p_0 z- \frac{1}{2}\left(m_i\Delta a z''+m_g g z\right)T\right]\right\}\\
    &\times\exp \left\{\frac{i}{\hbar}\left[p_0 - \frac{1}{2}\left(m_i\Delta a+m_g g\right)T\right]z'\right\}.
    \label{Eq.I8}
\end{split}
\end{align}

Next we choose $T$ such that
\begin{align}
    T\equiv \frac{2p_0}{m_i\Delta a+m_g g},
    \label{Eq:T}
\end{align}
which cancels the phase in the last exponent. We emphasize that this choice of $T$ is identical to the one obtained within classical mechanics discussed in the main text of our reply.

Now we perform the integration over $z'$ with the help of the identity
\begin{align}
    \int dz' \exp \left[\frac{i}{\hbar} \frac{m_i}{T} (z''-z)z' \right]=\frac{2\pi\hbar T}{m_i}\delta(z''-z),
\end{align}
which allows us to use again the definition, Eq. \eqref{Eq:T}, of $T$ to eliminate the phase factor in the second term in Eq. \eqref{Eq:I2}, leading us to the final result
\begin{align}
    I=e^{i \left(\phi_g -\phi_{\Delta a}\right)}.
\end{align}

Here we have used the normalization of the initial wave function $\psi$ and recalled the definition, Eq. \eqref{Eq:N}, of the prefactor of the propagator.

\end{appendix}

\end{document}